\documentclass{ws-ijbc}

\begin{document}

\markboth{Marius-F. Danca}{Approach of a class of discontinuous
dynamical systems of fractional order}

\title{Approach of a class of discontinuous dynamical systems of fractional order:
existence of the solutions}
\author{Marius-F. Danca\\Department of Mathematics and Computer Science, Avram Iancu University, \\Str. Ilie Macelaru, nr. 1A, 400380 Cluj-Napoca, Romania,\\and\\Romanian Institute of Science and Technology, \\Str. Ciresilor nr. 29, 400487 Cluj-Napoca, Romania}

\maketitle

\begin{abstract}
In this letter we are concerned with the possibility to approach the
existence of solutions to a class of discontinuous dynamical systems
of fractional order. In this purpose, the underlying initial value
problem is transformed into a fractional set-valued problem. Next,
the Cellina's Theorem is applied leading to a single-valued
continuous initial value problem of fractional order. The existence
of solutions is assured by a P\'{e}ano like theorem for ordinary
differential equations of fractional order.

\end{abstract}

\bigskip

Let us consider a general class of autonomous discontinuous
dynamical systems of fractional order modeled by the following
Initial Value Problem (IVP)
\begin{equation}
D_{\ast}^{q}x=f\,(x):=g\,(x)+A~s(x)\,,~\ x^{(k)}(0)=x_{0}^{(k)}%
\quad(k=0,1,\ldots,\lceil q\rceil-1),~t\in I=[0,\infty), \label{IVP
initiala}
\end{equation}
\noindent where$\ \,g:\mathbb{R}^{n}\rightarrow\mathbb{R}^{n}$ is a
function continuous with respect to the state variable, $A=\left(
a_{i,j}\right)  _{n\asymp n}$ a real constant matrix and $s~$is a
piecewise continuous function given by
\[
s(x)=\left(
\begin{array}
[c]{c}%
sgn(x_{1})\\
\vdots\\
sgn(x_{n})
\end{array}
\right)  .
\]
\noindent$q\in%
\mathbb{R} _{+}$ and $D_{\ast}^{q}$ is considered in this letter as
being the most utilized differential operator, the Caputo operator
of order
$q$ with starting point $0$, i.e. (see e.g. \cite{Podlubny})%
\[
D_{\ast}^{q}u(t)=\frac{1}{\Gamma(\lceil
q\rceil-q)}\int_{0}^{t}(t-\tau )^{\lceil q\rceil-q-1}u^{(\lceil
q\rceil)}(\tau){\mathrm{d}}\tau.
\]

\noindent $\Gamma:(0,\infty)\longrightarrow%
\mathbb{R} $ is the known Euler's Gamma function and
$\lceil\cdot\rceil$ denotes the ceiling function that rounds up to
the next integer. Thus, $D_{\ast}^{\lceil q\rceil}$ is the
conventional differential operator of order $\lceil
q\rceil\in\mathbb{N}$.

\begin{remark}
\noindent According to the standard mathematical theory \cite[\S
42]{stefan}, we are forced to give the initial conditions for the
IVP (\ref{IVP initiala}) using fractional derivatives of the
function $f,$ or these values are frequently not available. Also, it
may not even be clear what their physical meaning is. Therefore,
using the Caputo's suggested way, the initial conditions may be
specified in the classical way, as in IVP (\ref{IVP initiala}).
\end{remark}

\noindent For the sake of simplicity, we restrict ourselves to the
case important for the applications: $q\in({0,1)}$ (however the
considerations in this paper can be generalized to arbitrary
positive $q$). Therefore, we deal with the following form of IVP
(\ref{IVP initiala})
\begin{equation}
D_{\ast}^{q}x=f\,(x):=g\,(x)+A~s(x)\,,~\ x(0)=x_{0}^{{}},~t\in
I=[0,\infty), \label{IVP generala}
\end{equation}
\noindent where we have to specify just one condition since it is
easily seen that the number of initial conditions that one needs to
specify in order to obtain a unique solution is $\lceil q\rceil=1.$

\begin{remark}
\noindent The IVP (\ref{IVP initiala}) are enough general to include
the great generality of systems: for $q=1~$the systems modeled by
the IVP (\ref{IVP initiala}) are the known Filippov systems
\cite{Filippov}, while for non integer values of $q$ and
$A=O_{n\times n}$, the IVP (\ref{IVP initiala}) models dynamical
systems of fractional order.
\end{remark}

\noindent The main result of this letter is the following theorem

\begin{theorem}
\label{tyeorema}The (\ref{IVP initiala}) admits at least one solution.
\end{theorem}

\noindent For a better readability of the letter, the proof of this
theorem shall be given in several steps.

\bigskip

\noindent\emph{1. Filippov regularization of the right-hand side}

The existence and uniqueness of solutions to discontinuous IVPs are
essential for discontinuous dynamical systems, because due to the
right-hand discontinuity, classical solutions of IVP might not even
exist. To provide the existence, it is necessary to modify the
right-hand side of IVP (\ref{IVP generala}). For discontinuous
vector fields, existence and uniqueness of solutions is not
guaranteed in general, no matter what notion of solution is chosen.
Also, the classical notion of solution for ordinary differential
equations is too restrictive when considering discontinuous vector
fields. A possibility to compass this difficulty is to extend the
notion of differential equation to differential inclusion, problem
solved by Filippov using a generalized concept of solution. Thus,
the single valued discontinuous IVP is shifted to the following set-valued one%
\begin{equation}
D_{\ast}^{q}{x}\in F(x),~x(0)=x_{0},~\text{for almost all }t\in
I,\text{ } \label{DI}
\end{equation}
\noindent where $F: \mathbb{R}^{n}\Longrightarrow \mathbb{R}
^{n}~$is a set-valued vector function defined on the set of all subsets of $%
\mathbb{R} ^{n}$. One of simplest definition of $F$ is the following
convex form (implicitly used in most introductory references)
\begin{equation}
F(x)=\bigcap_{\varepsilon>0}\bigcap_{\mu\,(M)=0}\overline{conv}
\,f\,((x+\varepsilon\,B)\backslash\,M)\,, \label{multifunction}
\end{equation}
\smallskip where, \noindent$M$ is the set of discontinuity points of
\ $f$,$~~B$ the unit ball in $\mathbb{R}^{n}$, $\mu$ the Lebesgue
measure and $\ \overline{conv}$ the closed convex hull. At the
points where the function $\ f$ is continuous, $F(x)$ \ will consist
of one point, which is the value of $\ f$ \ at this point, i.e.
$F(x)=\{f(x)\}$. At the discontinuity points, the set$\ F(x)$ is
given by (\ref{multifunction}). Therefore, $F(x)$ is the convex hull
of values of \ $f$ $(x^{\ast}),~x^{\ast}\in M$, ignoring the
behavior on null sets. \noindent For example, the Filippov
regularization applied to the unidimensional \emph{sign} function
leads to the set-valued function
\[
Sgn\,(x)=\left\{
\begin{array}
[c]{ll}%
\left\{  -1\right\}  & x<0,\\
\left[  -1,\,1\right]  & x=0,\\
\left\{  +1\right\}  & x>0.
\end{array}
\right.
\]
\noindent and the right-hand side of the IVP (\ref{IVP generala})
becomes
\begin{equation}
F(x):=g\,(x)+A~S(x),~\noindent\text{with~~}S(x)=\left(  Sgn(x_{1}%
),\ldots,Sgn(x_{n})\right)  ^{T}. \label{membrul drept}%
\end{equation}
\noindent In order to justify the use of the Filippov regularization
in physical systems, $\varepsilon$ in (\ref{multifunction}) must be
small enough, so that the motion of the physical system can be
arbitrarily close to a certain solution of the differential
inclusion.

\bigskip

\noindent\emph{2. Continuous approximation of the right-hand side}

\noindent Next, $X$ and $Y$ denote metric spaces (e.g. $ \mathbb{R}
^{n}$ as in almost real applications).
\begin{definition}
A selection of a given set-valued function $F:X\Longrightarrow Y$ is
a function $h:X\longrightarrow Y$ satisfying
\[
\forall x\in X,~~h(x)\in F(x).
\]
\end{definition}
\begin{definition}
A set-valued function $F:X\longrightarrow Y$ is called upper
semicontinuous (u.s.c.) at $x\in X$ if for any neighborhood $V$ of
$F(x)$, there exists a neighborhood $U$ of $x$ such that
$F(x)\subset V$ for all $x\in V.$ $F$ is u.s.c. on $X$ if it is
u.s.c. on every point of $X$.
\end{definition}
\noindent For practical reasons, it is convenient to characterize a
set-valued map $F:X\Longrightarrow Y~$by its graph
\[
Graph(F)=\{\left(  x,y\right)  \in X\times Y~|~y\in F(x)\}.
\]
\begin{proposition}
\label{usc}The set-valued function $F$ defined by (\ref{multifunction}) is
u.s.c. with nonempty closed and convex values.
\begin{proof}
The proof can be found e.g. in \cite[p. 102]{Aubin1}.
\end{proof}
\end{proposition}

\begin{remark}
\label{remarca cu graf} Due to the symmetric interpretation of a
set-valued map as a graph (see e.g. \cite{Aubin2}) we shall say that
a set-valued map satisfies a property if and only if its graph
satisfies it. For instance, a set-valued map is said to be convex if
and only if its graph is a convex set.
\end{remark}

\noindent The following known theorem (Cellina's Theorem or
"Approximative Selection Theorem") will be a main tool used in the
proof of Theorem \ref{tyeorema}.

\begin{theorem}
\label{th Aubin}\cite{Aubin1,Aubin2} Let $F:X\longrightarrow Y$ be
upper u.s.c. set-valued with$\ Y$ a Banach space. If the values of
$F$ are nonempty and convex then, for every $\varepsilon>0$, there
exists a locally Lipschity function
$f_{\varepsilon}:X\longrightarrow Y$ such that
\[
Graph(f_{\varepsilon})\subset Graph(F)+\varepsilon B.
\]
\end{theorem}

\noindent The proof of Theorem \ref{th Aubin} is constructive (see
the mentioned reference) in that it provides a method to explicitly
construct the selection $f_{\varepsilon}$ with $\varepsilon$
parameter. This is an important advantage which allows to find
practical approximations in real applications.

\bigskip

\noindent\emph{3. Existence of the solutions for fractional
equations}

Let us consider the following IVP of fractional order with $q\in(0,1)$
\begin{equation}
D_{\ast}^{q}x=f(x),~x(0)=x_{0}, \label{ivp FRACT GEN}
\end{equation}

\noindent with $x\in \mathbb{R} ^{n}.~$The following existence
\smallskip result corresponds to the classical P\'{e}ano existence
theorem for first order equations

\begin{theorem}
\label{Kai th}\cite{Kai carte} Assume $f$ in IVP (\ref{ivp FRACT GEN}) is
continuous and bounded. Then there exists a solution to IVP
(\ref{ivp FRACT GEN}).
\end{theorem}

\begin{proof}
see \cite[Corollary 6.4 p. 92]{Kai carte}.
\end{proof}

\noindent Next we can give the proof for Theorem \ref{tyeorema}
\begin{proof}
Applying Filippov regularization, the right hand side of (\ref{IVP
initiala}) transforms into the set-valued function $F$ given by
(\ref{membrul drept}). $F~$is a convex u.s.c. (Proposition
\ref{usc}) and non-empty valued function (Remark \ref{remarca cu
graf}). Therefore, Theorem \ref{th Aubin} can be used and the IVP
(\ref{IVP initiala}) becomes a continuous IVP of fractional order to
which Theorem \ref{Kai th} applies and the proof is complete.
\end{proof}

\begin{remark}
i) Using the Fillipov regularization, the single valued IVP may be
considered as transforming into a set-valued IVP of fractional
order. However, for practical purposes, fractional single-valued
IVPs are more accessible by numerical point of view, since there are
several ways to approximate their solutions.\\
ii) The uniqueness for general fractional equations is treated in
\cite[\S 6.2]{Kai carte} the underlying theorem corresponding to the
well-known Picard- Lindel\"{o}f Theorem for equations of integer
order and is based on Lipschitz continuity. However, for our case of
systems, the uniqueness is checked since from Theorem \ref{th Aubin}
under some a priori boundedness, the approximate selection is
locally Lipschitz continuous.
\end{remark}

\begin{figure}[h]
\begin{center}
\psfig{file=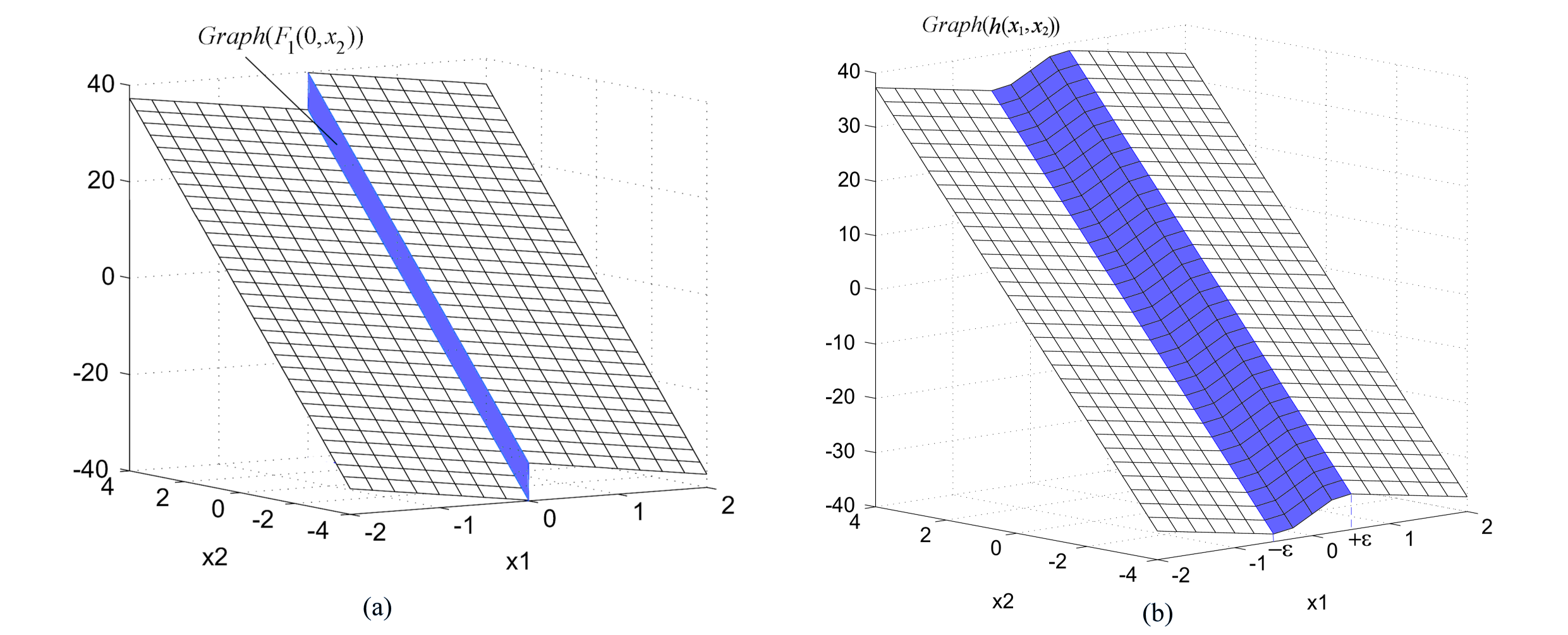, width=1\linewidth}
\end{center}
\caption{a) Graph of $F_1(0,x_2)$; b) Approximation of $F_1$.}
\label{fig0}
\end{figure}

\noindent For example, let us consider the following fractional
discontinuous variant of the Chua's system
\begin{equation}
\begin{array}{l}
D_{\ast }^{q}x_{1}=-2.57x_{1}+9\,x_{2}+3.87\,\,sgn\,(x_{1}), \\
D_{\ast }^{q}x_{2}=x_{1}-x_{2}+x_{3}, \\
D_{\ast }^{q}x_{3}=-p\,x_{2},\,\label{chua disc}
\end{array}
\end{equation}
\noindent where $p$ is a real parameter. The graph of the first
component of the right-hand side of (\ref{chua disc}), the scalar
function $f_{1}\left(  x_{1},x_{2}\right)  =$ $-2.57x_{1}
+9\,x_{2}+3.87\,\,sgn\,(x_{1})$, is plotted in Fig. \ref{fig0}(a).
The convex hull of $f_{1}\left(  0,x_{2}\right)  $ is the dashed
region and represents the graph
of the set-valued function $F_{1}(0,x_{2}).$ The underlying set-valued IVP is%
\begin{equation}
\begin{array}
[c]{l}%
D_{\ast }^{q}x_{1}\in-2.57x_{1}+9\,x_{2}+3.87\,\,Sgn\,(x_{1}),\\
D_{\ast }^{q}x_{2}=x_{1}-x_{2}+x_{3},\\
D_{\ast }^{q}x_{3}=-p\,x_{2}.\ \label{chua disc2}
\end{array}
\end{equation}
\noindent A possible approximation (selection) of $F_{1}$ in the
neighborhood of $(0,x_{2})$, for $x_{2}\in \mathbb{R}$, can be for
example a smooth cubic polynomial function $h\left(
x_{1},x_{2}\right) =ax_{1}^{3}+bx_{1}^{2}+cx_{1}^{{}}+9x_{2}$
\cite{danaca si codr} where $a,~b,~c$ and$~d$ have to be determined
from the continuity and differentiability conditions in $x_1=\
\pm\varepsilon$ (Fig. \ref{fig0}(b).

\noindent Next, the system can be either numerically integrated
using e.g. the fractional Adams--Bashforth--Moulton method discussed
in \cite{Diethleme et al}, or approximated using some
frequency-domain method\emph{ (}see e.g. \cite{Charef}). More
approximations ways for fractional operators can be found e.g. in
\cite{Vinagre}.

\end{document}